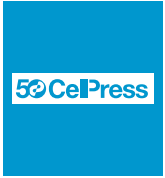
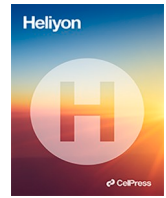

Research article

# Tumor microenvironment (Part I): Tissue integrity in a rat model of peripheral neural cancer

Ahmad Maqboul [a,b,*], Bakheet Elsadek [b]

[a] *Department of Anesthesiology and Operative Intensive Care Medicine, Charité–School of Medicine Berlin, Corporate Member of Freie Universität Berlin and Humboldt Universität zu Berlin, Berlin, Germany*
[b] *Department of Biochemistry and Molecular Biology, College of Pharmacy, Al-Azhar University, Asyût, Egypt*

ABSTRACT

ICAM–1 (intercellular adhesion molecule 1) and MPZ (myelin protein zero) are thought to be a factor in the integrity of nerve tissues. In this report, we attempted to trace the expression of ICAM–1, responsible for cell-to-cell adhesion, and of MPZ, the main constituent of myelin sheath, in malignant tissues of the sciatic nerve (SN) in inbred male Copenhagen rats. AT–1 Cells (anaplastic tumor 1) were injected in the perineurial sheath, and tissues of the SNs were collected after 7, 14 and 21 days and compared to a sham-operated group of rats (n = 6 each). Tissues were sectioned and histologically examined, under light microscope, and stained for measuring the immunoreactivity of ICAM–1 and MPZ under laser scanning microscope. The cancer model was established, and the tumor growth was confirmed. ICAM–1 showed severe decreases, proportional to the growing anaplastic cells, as compared to the sham group. MPZ revealed, however, a distinct defensive pattern before substantially decreasing in a comparison with sham. These results support the notion that malignancies damage peripheral nerves and cause severe axonal injury and loss of neuronal integrity, and clearly define the role of ICAM–1 and MPZ in safeguarding the nerve tissues.

## 1. Introduction

The neuron doctrine has established that nerve cells construct the nervous system individually forming syncytial networks. Each neuron is designed with bundles of axons, and each axon is surrounded by myelin sheaths, with its distinctive Schwann cells [1]. In our project, we are trying to explore the patterns of expressing different classes of proteins participating in neuronal invasion, nerve damage, and regeneration and immune response within and distant from the complex tumor microenvironment (TME). These classes include neurotrophins, cytokines, cytochromes and epoxyeicosanoids, apoptotic, oncogenic and transcription factors, glia and astrocyte markers, and ion channels. Therefore, we first transplanted AT–1 (anaplastic tumor 1) cells in Copenhagen rats, which enabled us to establish the model of induced peripheral neural cancer, and to generate tissue samples constantly and reproducibly from the malignant lumps. These tissues were used to deeply explore the TME and to investigate the regulation of the underlying proteins responsible for cell adhesion, myelination, and neural growth and survival. Here, in the context of tissue integrity, we have examined ICAM–1 (intercellular adhesion molecule 1; CD "cluster of differentiation" 54) and MPZ or P0 (myelin protein zero). ICAM–1, on one hand, endorses, as its name indicates, adhesion of the cells to each other and to the extracellular matrix, organizes intracellular responses, and recruits activated leukocytes in cases of inflammation [2]. MPZ, on the other hand, is the main constituent and the most abundant protein of the myelin sheaths. Its functions include myelination, insulation and protection of the axons, and propagation of the nerve impulses [3]. In earlier studies on animal peripheral nerves, malignant cells were inoculated in the locality of the sciatic nerve (SN) [4,5]. Here we bypassed the effect of the surrounding tissues by directly injecting the anaplastic cells inside the envelope surrounding the SN. This permitted a more objectivity in investigating different neurotrophic factors and receptors arising from the continuously growing tumor.

## 2. Materials and methods

*Copenhagen rats (COP/CrCrl)*: Copenhagen rats have been enrolled due to their MHC (major histocompatibility complex) of the RT1$^{av1}$ haplotype which permitted malignant cells with a hundred-percent-growing rate [6]. To exclude any possible effects of sex hormones, and although the AT-1 cells are estrogen- and androgen-receptors negative, we preferred to recruit only inbred male rats [7,8]. Including female rats, however, might have had no increased variability according to a recent report [9]. The animals have been purchased from Charles River Laboratories International, Inc., Germany, through the Research Institutes for Experimental Medicine (FEM), Charité-School of Medicine Berlin. They have been sent (n = 6 for each group) approximately weighing 250 g and have been bred with numbers from 2 to 4 rats in each cage. Nutritional and water supply have been continually provided with a light source of two light and dark cycles a day. The experiments were performed in a random manner, and the allocation codes have been kept out of the examiner's hands throughout the experiments' times. The animals were to be euthanized upon the detection of any indicators of discomfort or distress. For this goal, an isolated space, with a transparent compartment supplied with 1.8–5.3 L/min $CO_2$ (around twenty percent flow rate), was set up [10]. However, none of the animals, after the optimization of the model, were excluded from the experiments.

*Cancer cell line and antibodies*: AT-1 Cells, originating from Dunning R3327, are described by being non-metastasizing with an incredibly fast progression in the absence of the androgens hormones [8]. Its main source, i.e. Dunning R3327, was first detected as a naturally occurring tumor in an old male of Copenhagen rat. It was seen at a specimen from the fifty fourth generation of its previous culture 2331 [6,11,12]. AT–1 Cell line was supplied from ECACC, UK (Catalogue number: 94101449). The anti-ICAM–1 antibody (mouse; Catalogue number: ab2213) was purchased from Abcam plc, Cambridge, UK and the anti–MPZ (chicken; Catalogue number: AB9352) from Millipore Corporation, MA, USA. Alexa Fluor® 594 (H + L) donkey anti-mouse IgG (Catalogue number: A-21203), and Alexa Fluor® 488 goat anti-chicken IgY (Catalogue number: A-11039) were bought from Thermo Fisher Scientific™ GmbH, Germany.

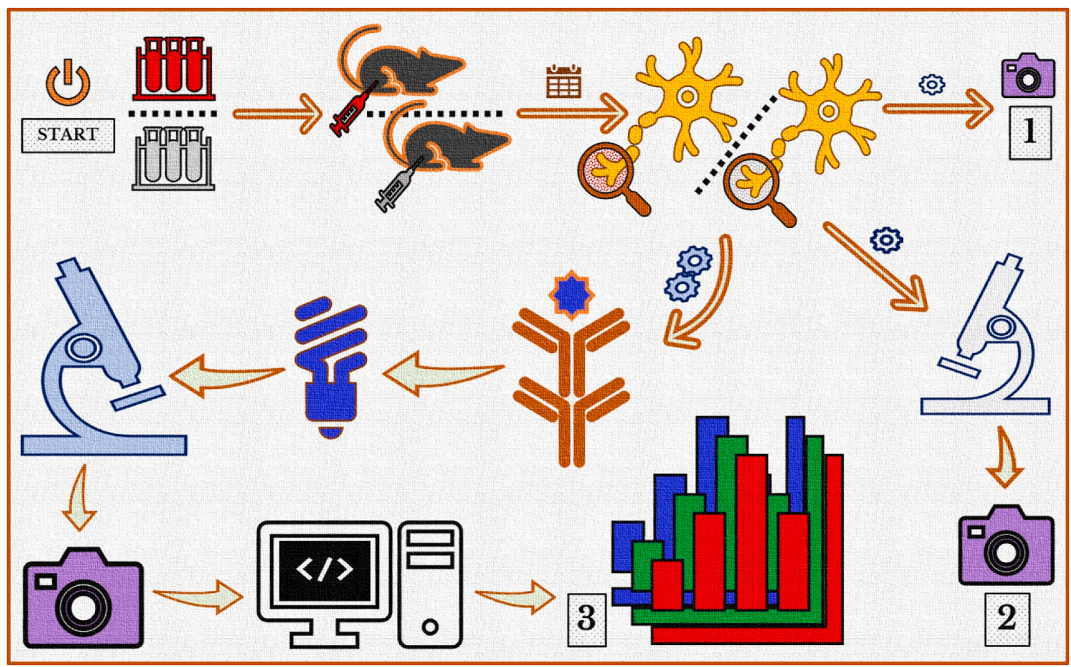

**Fig. 1.** Schematic representation of the experimental design. AT-1 Cells were injected within the SNs' perineurial sheath of Copenhagen rats. Different groups were recruited for each timepoint. A sham-operated group was also recruited, in all steps, for comparison purposes. Three paths were taken to perform these experiments. The morphological changes of the extracted SNs were photographed (**1**). Tissues of the SNs were sliced, stained by H&E, and visualized under light microscope (**2**). For the third path, SN-tissues have been taken away using a different protocol, incubated with the primary and then the fluorochrome-tagged-secondary antibody. Fluorescence captured by confocal laser microscope has been further processed, and the measured intensities were statistically analyzed (**3**).

*Cancer cells inoculation*: AT–1 Cells were flourished in RPMI 1640, with the addition of 2 mM L-glutamine, 250 nM dexamethasone and 10% foetal bovine serum, kept at five percent $CO_2$ and thirty seven Celsius degrees, and counted with a Hemacytometer slide (Bright-Line™) by means of a simple microscope. In order to perform the surgery, the animals were hypnotized by an $O_2$ inhalation with isoflurane. The SN was uncovered by making a small opening between the gluteal muscle and the biceps after sterilization of the right back limb using iodine and alcohol. Cell culture was prepared in PBS (phosphate buffered saline) with a pH adjusted to 7.4 ($1.0 \times 10^6$ cells in 10 μL PBS) and gradually injected beneath the nerve covering by the aid of a Hamilton® syringe (GASTIGHT® 1702LT Series, 25 μL; Sigma-Aldrich Chemie GmbH, Germany). Throughout the initial seven days, animals showed an unprompted ache with a changing grade of inflammatory reactions in the place of surgery, which is attributed, upon euthanasia, to a very apparent tumor infiltration outside the nerve and in compact adherence to the internal tissues. Therefore, we reduced the amount to $0.5 \times 10^6$ cells per 10 μL PBS. The progression of the tumor was declined with neither an incursion to the adjacent tissues nor exaggerated adverse events. This number of cells was used for all cancer groups (*cf.* pancreatic cells in a concentration of $1.0 \times 10^5$ per microliter buffer [13]). To exclude the effects that may arise from the surgical procedure we injected a group of animals with the vehicle, i.e. PBS, and used it as a control. The post-surgical analgesic used, metamizole sodium, was injected intraperitoneal, and also inserted in bottles of water for the following 72 h. Repetitive inspections were conducted every 24 h to confirm regular movements and deficient signs of discomfort, e.g. bending or twisting, lacked consumption, and neglected cleaning actions (Fig. 1).

*Tissue collection*: Tissues were collected from sham-operated and cancer-inoculated animals, after 7, 14, and 21 days, in two separate procedures. The first for the investigation of morphology and light microscopy and the second for the study of immunofluorescence (Fig. 1).

*For morphology and light microscopy*: Rats were anaesthetized with isoflurane, and the SNs, around 2.5 cm long, were extracted into Eppendorf® Tubes (Eppendorf AG, Germany), instantly submerged within liquid nitrogen, placed in a −80 °C refrigerator, and further checked for the development of neoplasms and the continuity of swelling. One to one-and-half centimeters of tumorous tissue were cut from the SNs. Nerve tissues were sliced into longitudinal sections. The prepared slides were histologically visualized by means of hematoxylin and eosin (H&E) and an inverted light microscope (Axiovert 25, Carl Zeiss, Germany) assembled with a cooled charge-coupled device (CCD) camera (AxioCam HRc).

*For immunofluorescence staining*: Rats were anaesthetized with isoflurane and a 100 mL PBS solution, with an adjusted pH of 7.4, was diffused *via* the heart throughout the circulation, followed by 300 mL of a four percent-paraformaldehyde solution in 0.1 M buffer (w/v). The SNs, around 2.5 cm long, were collected in Eppendorf® Tubes, also in paraformaldehyde, and left at room temperature for 90 min. The nerve tissues were rinsed with the buffer and kept in a ten percent sucrose solution in buffer at a 4 °C refrigerator to the following morning. Tissues were, then, immersed within a "optimal-cutting-temperature" material and retained at −20 °C.

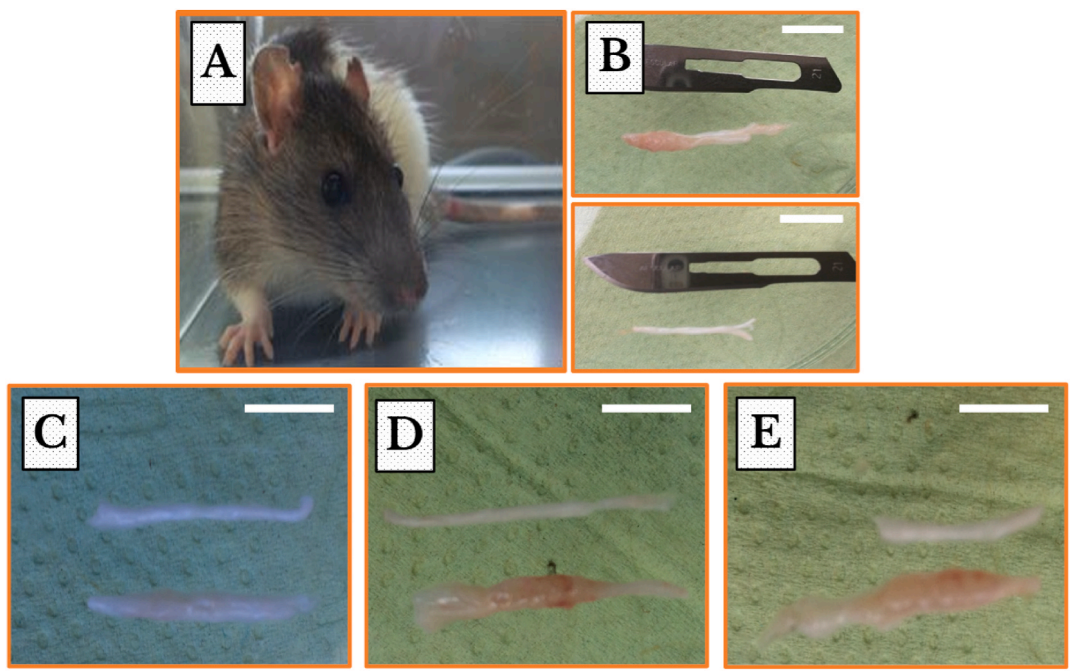

**Fig. 2.** A Copenhagen rat (*COP/CrCrl*) and morphological changes in the extracted SNs immediately captured after surgery. (**A**) A Copenhagen rat with its characteristic white color and brown hood. (**B**) Photos of exposed SNs, around 2.5 cm long, from both ipsilateral (*upper photo*) and contralateral (*lower photo*) sides of the same animal. (**C**–**E**) Photos of exposed SNs from sham-operated (*upper*) and cancer injected (*lower*) animals. The continuous proliferation of the malignant cells within the SNs in the cancer-inoculated rats after seven (**C**), fourteen (**D**), and twenty one (**E**) days are observed. Scale bars = 1 cm. (For interpretation of the references to color in this figure legend, the reader is referred to the Web version of this article.)

Approximately one centimeter of tumorous tissue was cut from the SNs. Tissue cubes were sliced into sections from five to 7 μm thickness by means of a cryostat (CryoStar™ NX70, Thermo Fisher Scientific™ GmbH, Germany). Tissue slides were incubated for 1 h with a PBS solution containing Triton X-100 (0.3 percent), BSA (bovine serum albumin, 1 percent), horse serum (5 percent) and donkey serum (5 percent). Slices were attached to slides covered with gelatin and kept at 25 °C with the intended antibodies for the next morning. Tissue slides were rinsed and retained with the secondary antibodies labeled with Alexa Fluor® 488 or Alexa Fluor® 594. Tissues were visualized by means of a confocal microscope (Zeiss® LSM 510, Germany). To maximize the signal and reduce background and noise, we used photo-stable fluorophores of high-quantum yield, a CCD camera, clean and close coverslips, and a minimally fluorescent mounting medium.

***Software-aided image analysis***: Fiji (ImageJ) is the program used in this study because of its capabilities of automated unbiased image analysis (ImageJ User Guide 1.46r) [14,15]. PIVs (pixel intensity values) in the ROIs (regions of interest) were estimated by transforming the captured fluorescent micrographs from a LSM (laser scanning microscope)- to a TIF (tagged image file)-type, programmed scaling the eight-bit photos, producing binaries, and subtracting the noises surrounding the ROIs. PIVs were generated and accumulated in a separate file of the CSV (comma-separated values) type for statistical analysis (see **Supplementary File**).

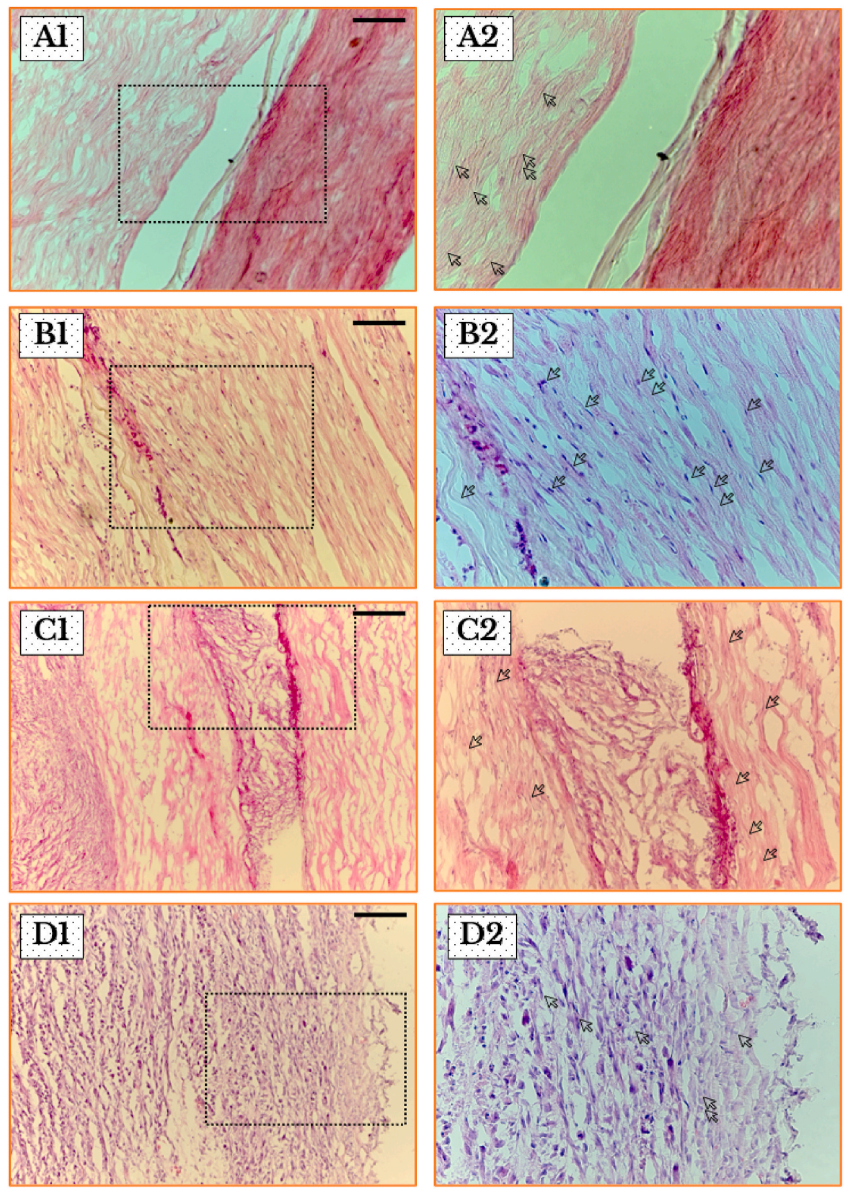

**Fig. 3.** Histopathological examination of longitudinal sections of SN stained by hematoxylin and eosin. (**A1** and **A2**) A light micrograph from sham-operated animal tissues (*left*-hand column, × 20) and its magnified insert (*right*-hand column, × 40) reveal smooth bundles and a curvy shape of fibers which portray regular SNs. (**B1**–**D1** and **B2**–**D2**) SN Tissues from cancer-injected animals, on days seven, fourteen and twenty one, respectively, display an infiltration of malignant cells with a gradual enlargement of Schwann cells nuclei and a degradation of nerve fibers. Scale bars = 100 μm.

*Statistical data analysis*: GraphPad Prism® (GraphPad Software, Inc., USA) was implemented in statistical assays. Data were presented as [means (standard errors)] and decimals were rounded to two significant digits [16]. To compare between the studied groups, PIVs were analyzed by one-way ANOVA (analysis of variances) and the Dunnett's post *hoc* test. If PIVs were not normally distributed, the variations of the mean ranks were assayed by the Kruskal-Wallis and the Dunn's tests. Statistical results of $P < 0.05$ were concluded as considerable.

## 3. Results

*SNs' Morphology*: We evaluated the macroscopical and microscopical changes, as a result of AT-1 cells injection, in the SNs. Tumors proliferated in the ipsilateral and not in the contralateral SNs with observed gradual thickening and weight gain. Tumor invasion also was not encountered in the sham-operated group (Fig. 2 (**A – E**)).

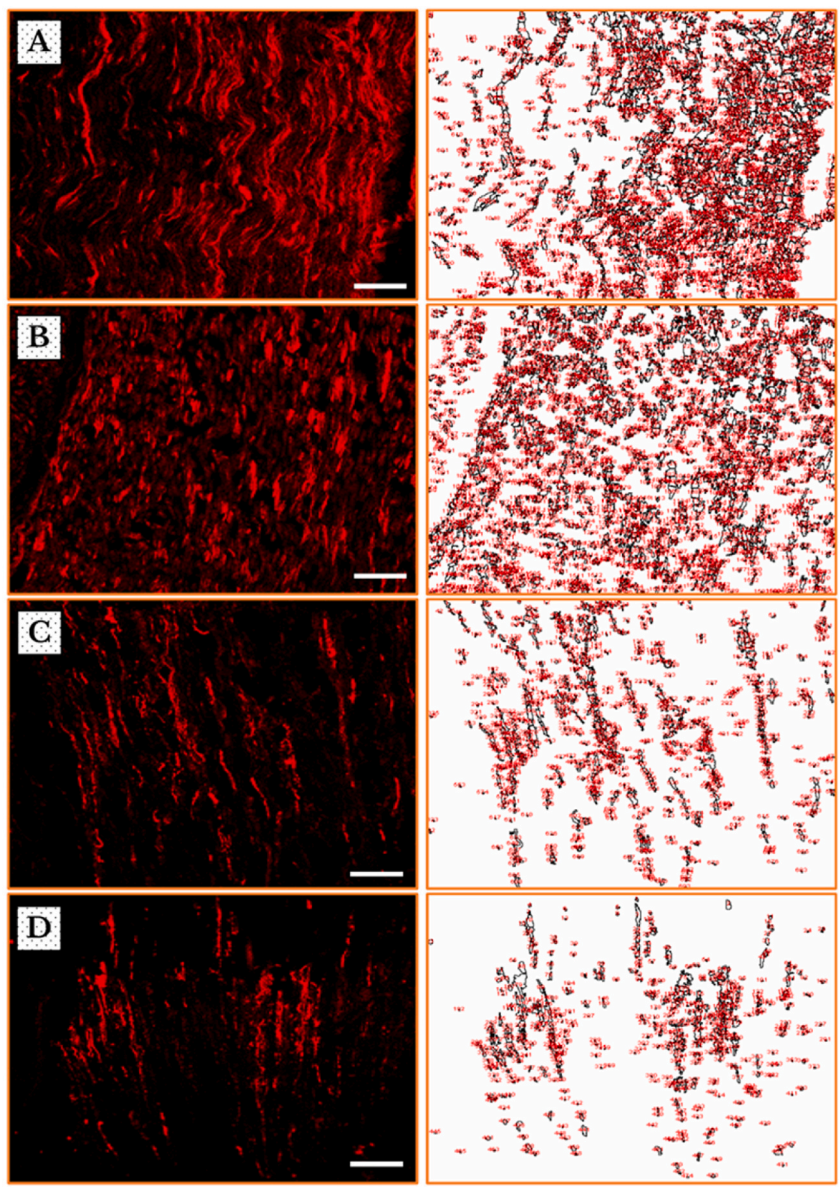

**Fig. 4.** Immunofluorescence images of ICAM–1 in the neurons [on the *left* side] and the corresponding regions of interests identified by ImageJ software [on the *right* side] in sham (**A**) and cancer-induced animals after 7 (**B**), 14 (**C**) and 21 (**D**) days of AT–1 cells inoculation in SN of Copenhagen rats. The images show a gradual loss of immunoreactivity, which is represented by the staining of the whole nerve fibers, from one (**B**) to two (**C**) and then three (**D**) weeks of cancer invasion as compared to sham (**A**). Scale bars = 50 μm.

*Histology of tumor tissues*: SNs have been checked for signs of cancer development, tissue disintegration, and infiltration of immune cells. Micrographs of the SNs in sham-injected group showed ordinary fascicles with a distinctive wavy shape and bar-like fibers. However, the fibrocytes and the Schwann cells' nuclei revealed no obvious distinction. In cancer-inoculated rats, SNs were infiltrated by mononuclear cells. In addition, the gradual growth of the tumor from day 7 to day 21 was indicated by loss of integrity (Fig. 3 (**A** – **D**)).

*Decreased expression of ICAM–1*: The visualization of the fluorescent micrographs, on the *left*-hand column, shows a decline of ICAM–1 positive immunoreactive (+IR) nerve fibers after 7 days of cancer cells inoculation, followed by a dramatic decrease to the halfway on day 14, which tries to regenerate on day 21, as compared to sham (Fig. **4** (**A**–**D**)).

The corresponding photos, on the *right*-hand column, clearly represent the difference in fluorescence, i.e. the numbers and areas identified as ROIs, between the sham-operated group in a comparison to the one-week group of tumor cells' injection, and further to the two- and three-weeks groups (Fig. **4** (**A**–**D**)).

These observations were confirmed by measuring the PIVs of ICAM–1 +IR nerve fibers showing a decrease on the seventh day [95.4 % (1.1)], severely going to a point of an equidistance after 14 [66.8 % (0.79)] and 21 [79.2 % (2.8)] days as compared to sham [100 % (1.6)]. The difference between the values after 14 [66.8 % (0.79)] and 21 [79.2 % (2.8)] days are also significant (Fig. 5).

*Modified expression of MPZ*: The immunofluorescent photos of MPZ + IR fibers behaved, however, in a different way. They remarkably showed a greater amount of MPZ + IR fibers after one week, which became substantially fewer after the second and third weeks. The results displayed a substantial rise in the animal group on the 21-days in a comparison with that 14-days one. The corresponding photos, on the *right*-hand column, show the increased regulation in fluorescence in the one-week group of tumor cells' injection as compared to the sham group, followed by severe decreases in the two- and three-weeks groups. These visual identifications were in an agreement with the assessed PIVs as exhibited from the corresponding ROIs in the photos on the *right*-hand column (Fig. **6** (**A**–**D**)).

The fluorescence of the MPZ + IR fibers in the one-week group of tumor growth has been very intensely increased on day 7 [124 % (1.9)] then decreased for the most on days 14 [72.8 % (0.7)] and 21 [72.2 % (1.6)] as compared to sham [100 % (1.9)]. However, there was no change in the intensity values between the groups of the second- and third week of cancer injection. (Fig. 7).

## 4. Discussion

*The TME in neural cancer*: Models of perineural cancer invasion were previously induced through the injection of invading malignant cells in the surroundings of that peripheral nerve [4,5]. Limitations, nevertheless, resulting from existent immune cells may arise. Therefore, and in order to avoid this obstacle, we have designed the study in a way that cancer cells were injected *in situ* giving an intimate contact to nerve fibers.

*Disintegration and demyelination of the nerve fibers, damage of the axons, and occupation by the invading anaplastic cells.* In the current study, photos represent an evidence for the continuous development of malignant lumps in the SN (Fig. 2). Light microscope photos revealed a penetration of mononuclear cells within the nerve bundles, and clearly identified internal neural structures and verified the degradation of nerve tissues characterized by gradual disintegration of the structural arrangement of the fascicles, thinning of the fibers, and damage and persistent dystrophy of the axons. The current study confirmed the occurrence of lesions in the myelin sheaths exposing the axons leading to their damage and the subsequent replacement of the nerve tissue with a growing lump of tumor cells.

*Automated determination of ROIs and measurement of pixel intensity values*: The novelty in the current report lie in the application of an automated method for measurement of pixel intensity values, for regions of interest (ROIs) in immunofluorescent micrographs, reflecting the expression of the studied proteins, ICAM-1 and MPZ, in the TME. The automation and the precision in determining the ROIs, which is traceable from the micrographs and their corresponding output photos, have given substantially accurate and time-saving measurements when compared to other traditional methods (Figs. 4 and 6). The lack of bias and vagaries is, above all, of a greater importance.

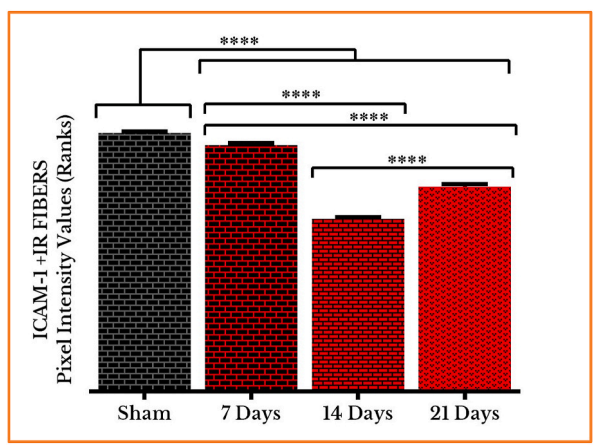

**Fig. 5.** Quantifying PIVs of ICAM–1 +IR fibers in SNs of vehicle- and cancer-injected groups (on days 7, 14 and 21). ICAM–1 positive immunoreactive nerve fibers decreased meaningfully on day 7, then a severely on days 14 and 21. (****) $P < 0.0001$.

***Loss of ICAM−1's adhesive function in rat's model of peripheral neural cancer.*** The continual decrease in ICAM−1's production, a protein distinguished by Rothlein et al., in 1986 [17], as compared to sham, confirms its main function in adhesion, and its failure to maintain this role once its stores were depleted and the cell machinery was disturbed. The inoculation of tumorous cells, in our *in vivo* cancer model, causes the recruitment of tissue-resident macrophages and other circulating inflammatory mediators to the invasion site, finally leading to a decrease in ICAM−1's production (*cf*. different responses in mammalian cell lines) [18–20]. This may give our model an advantage over other studies on cell lines, as responses to different stimuli, in different cell types or tissues and even in different reactive regions within the same tissue, cannot be identical. ICAM-1's increased expression on day 21, as indicated by higher PIVs of +IR nerve fibers, which is significant from day 14, may, however, be understood as a trial from these tissues to regenerate and to resist the loss of adhesiveness caused by the invading malignant cells. This rise is still a decline when compared to the higher levels of the sham and, even, of the 7 days' group.

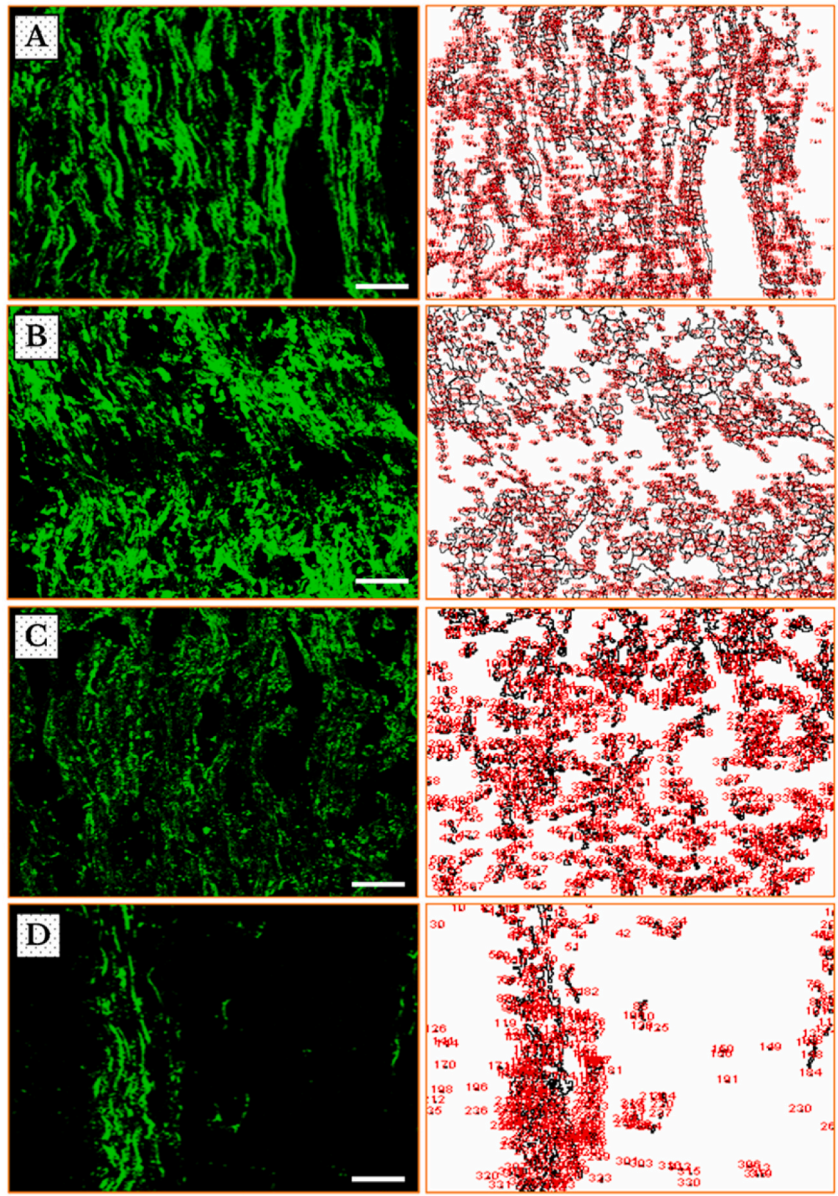

**Fig. 6.** Immunofluorescence micrographs of MPZ in the neurons [on the *left* side] and the corresponding regions of interests identified by ImageJ software [on the *right* side] in sham (**A**) and cancer-induced animals after 7 (**B**), 14 (**C**) and 21 (**D**) days of anaplastic tumor cells inoculation in SNs of Copenhagen rats. The images show an increase of immunoreactivity, which is represented by the more bright staining of the whole nerve fibers after one week (**B**), followed by severe decreases after two (**C**) and three (**D**) weeks of cancer invasion as compared to sham (**A**). Scale bars = 50 μm.

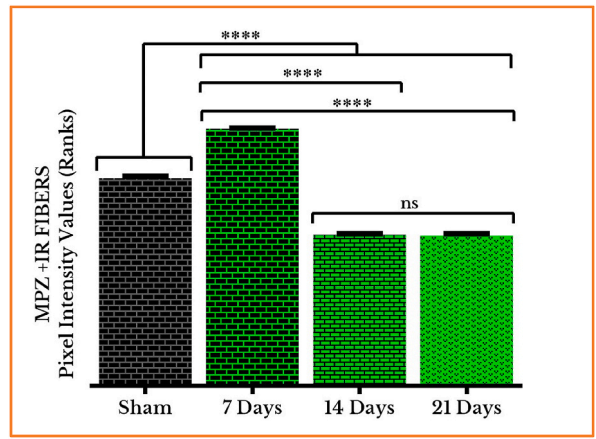

**Fig. 7.** Quantifying PIVs of MPZ + IR fibers in control and cancer-injected rats' SNs (on days seven, fourteen, and twenty one). MPZ positive immunoreactive nerve fibers showed a substantial rise after one week, and then severe declines at the end of the second and third weeks. There was non significant (ns) change between groups of the 14 and the 21 days. (****) $P < 0.0001$.

*Similar expression profiles of ICAM-1 in other tissues and cells.* Our experimental model on ICAM-1 in peripheral neural cancer is consistent with studies, on other types of tissues and cells, showing its declined expression. In breast cancer's tissues and cells, ICAM-1 exhibited declined levels as compared to normal breast epithelia or benign breast cells. ICAM-1's production have been stimulated by tumor necrosis factor alpha (TNFα). This was not the case with colony stimulating factor (CSF), interleukins 2 and 6 (IL-2 and IL-6), and interferons alpha and gamma (IFNα and IFNγ) [21]. It was also reported that ICAM-1's production have been decreased in stomach cancer which is correlated to lymphatic metastasis [22]. The decreased frequency of ICAM-1-expressing malignant cells was correlated to their metastasizing capabilities [23]. Another study has shown that ICAM-1 is exclusively expressed in aggressive tumor cells linked to the development of tertiary lymphoid structures (TLS) such as triple negative breast cancer (TNBC) and HER2 (human epidermal growth factor receptor 2), and that an encouraged ICAM-1's expression, in a number of cell lines, was observed by the pro-inflammatory cytokines TNFα, IL-1β (interleukin 1β) and IFNγ [24].

*Different expression profiles of ICAM-1 in other tissues and cells.* In contrast to our results, ICAM-1 was up-regulated in the following models. ICAM-1's production was enhanced in glioblastoma through the incubation of TAMs (tumor associated macrophages) in a hypoxic conditions with the addition of a HIF (hypoxia-inducible factor)-stabilizing drug. This increased expression has been concluded to promote tumorigenesis and glioblastoma aggressiveness. In context of this conclusion, and as a therapeutic approach, tumor cells were intracranially injected into ICAM-1 deficient mice leading to a longer survival with a lower overall tumor volume comparet to the wild type [25]. ICAM-1 was also overexpressed in bevacizumab-resistant glioma stem cells in hypoxic circumstances. The hypoxia-induced overexpression of ICAM-1 was linked to the up-regulation of p-STAT3 (phosphorylated signal transducer and activator of transcription). Overwhelming *ICAM-1* restrained cancer invasiveness in different models [26]. ICAM-1 was also overexpressed in TNBC. Guo and colleagues have designed a model of a MRI (magnetic resonance imaging)-probe depending on ICAM-1 antibody-coupled particles leading to an improved ICAM-1's production [27].

**MPZ's defensive role in regenerating the myelin sheaths and protecting the neural axons.** The myelinating protein showed an early protective response trying to insulate the nerve fibers and minimize axonal exposure. This trial delayed the damage for one week, which didn't succeed after two weeks, with the continuous growth of the aggressive malignant cells, to protect the axons. This conclusion is in line with studies referring to MPZ as the most abundant protein in myelin guaranteeing tissue integrity [28]. The immunostaining of MPZ reveals the characteristic shape of the nerve fibers which changes in agreement with its increased then decreased expression with the development of malignant tumors.

*MPZ's defferential expression patterns.* It is well established that mature Schwann cells express the neuronal markers, MPZ, and others, insulating and protecting large nerve axons with myelin sheaths [29]. Transgenic mice expressing active Fyn (a tyrosine kinase) displayed an elevated MPZ's production as compared to control [30]. Whole genome microarray analyses for samples of urinary bladder cancer and controls identified *MPZ*, with other 16, as common differentially expressed genes [31]. The demyelination of RSC96 (rat Schwann cells) characterized by a decreased expression of MPZ was also caused by the platinum-based chemotherapeutics cisplatin and carboplatin but not oxaliplatin [32]. A CGH (comparative genomic hybridization) analysis for a 6-member family from three sequential generations, diagnosed with demyelination and neural inflammation, revealed *MPZ*'s enhanced dosage. Therefore, tissue integrity of myelin was correlated to a significant *MPZ*'s dosage function and an elevated manufacturing of MPZ mRNA [33].

**Loss of tissue integrity and failure of neural regeneration.** Our report confirms that nerve tissues have been degenerated, and axons severely damaged, as a result of the progressive tumor lesions. These lesions are irreversible. It was noted, on one side, that the expression of ICAM−1 was down-regulated without presenting any other role in opposing the growth of the AT−1 cells. It was observed, on the other side, that the expression of MPZ has been increased as a trial to rebuild the myelin sheath to protect the axon and allow nerves to regenerate. As the trial did not succeed, the increased expression of MPZ showed a severe decay. These all indicate a permanent proliferation and an eternal replication of cancer cells, and a failed neural regeneration.

*Prospective studies.* The late rise of the expression level of ICAM-1 requires more examination as a target protein with potential anti-tumorigenesis and anti-proliferative therapeutic magnitudes. An investigation of the molecular mechanisms underlying the link between the declined expressions of ICAM-1 and MPZ from one side, and the regulations of related cytokines, and apoptotic and oncogenic factors, needs also to be conducted within and distant from this TME.

## Ethics statement

The project was approved by the State Office for Health and Social Affairs (LAGeSo, Berlin, Germany) according to the guidelines of the Charité - School of Medicine Berlin (Project code: G 0314/13) and in agreement with ARRIVE 2.0 guidelines (https://arriveguidelines.org/).

## Funding

This work was funded by the *Cultural Affairs and Missions Sector, Ministry of Higher Education* of Egypt, for two years; and the *Foundation of Prof. K.H. René Koczorek* of Germany, grant No.: IA89838780, for one year. The funders had no role in study design, data collection and analysis, decision to publish, or preparation of the manuscript.

## Data availability

The raw data are submitted as a **Supplementary File**. Further enquiries can be directed to the corresponding author.

## CRediT authorship contribution statement

**Ahmad Maqboul:** Writing – original draft, Visualization, Validation, Software, Methodology, Investigation, Funding acquisition, Formal analysis, Data curation, Conceptualization. **Bakheet Elsadek:** Writing – review & editing, Supervision, Project administration, Funding acquisition, Conceptualization. The research question of this study is a part of the PhD proposal registered by A. M. at the Charité - School of Medicine, Berlin, Germany.

## Declaration of competing interest

The authors declare that they have no known competing financial interests or personal relationships that could have appeared to influence the work reported in this paper.

## Acknowledgement

Time had taken a highly expensive toll in health, in trust, in humanity, and in money. The first author acknowledges the great support of his family members during this time. The authors thank their colleagues, from Charité - School of Medicine, Berlin, Germany, and Al-Azhar University, Asyût, Egypt, who contributed to this work.

## Appendix A. Supplementary data

Supplementary data to this article can be found online at https://doi.org/10.1016/j.heliyon.2024.e33932.